\documentclass[prd,aps,twocolumn,a4paper,floatfix,showpacs]{revtex4}

\usepackage{graphicx,psfrag}
\usepackage{mathrsfs}

\usepackage{multirow}
\usepackage{hyperref}
\usepackage{comment}
\usepackage{color}

\newcommand{\be}{\begin{equation}}
\newcommand{\ee}{\end{equation}}
\newcommand{\bea}{\begin{eqnarray}}
\newcommand{\eea}{\end{eqnarray}}
\newcommand{\bel}{\begin{align}}
\newcommand{\eel}{\end{align}}

\def\p{\partial}

\begin{document}

\title{Simulations of rotating neutron star collapse with the puncture gauge:\\
  End state and gravitational waveforms}

\author{Tim Dietrich$^1$, Sebastiano Bernuzzi$^{2,3}$}
\address{${}^1$Theoretical Physics Institute, University of Jena,
07743 Jena, Germany} 
\address{${}^2$TAPIR, California Institute of Technology, 1200 East
California Boulevard, Pasadena, California 91125, USA} 
\address{${}^3$DiFeST, University of Parma,  I-43124
Parma, Italy} 

\date{\today}

\begin{abstract} 
  We reexamine the gravitational collapse of rotating neutron stars to black
  holes by new 3+1 numerical relativity simulations employing the Z4c
  formulation of Einstein equations, the moving puncture gauge
  conditions, and a conservative mesh refinement scheme for the
  general relativistic hydrodynamics.   
  The end state of the collapse is compared to the vacuum spacetime
  resulting from the evolution of spinning puncture initial data.
  Using a local analysis for the metric fields, we demonstrate that
  the two spacetimes actually agree.
  Gravitational waveforms are analyzed in some detail. We connect the
  emission of radiation to the collapse dynamics using simplified
  spacetime diagrams, and discuss the similarity of the waveform
  structure with the one of black hole perturbation theory. 
\end{abstract}


\maketitle

\section{Introduction}

A fundamental problem in general and numerical relativity is the
simulation of the gravitational collapse of a rotating neutron star to a black
hole. The problem is of theoretical and astrophysical relevance,
and it has been studied in some detail by means of two-dimensional axisymmetric 
simulations~\cite{Stark:1985da,Nakamura:1987zz,Shibata:2003iy,Duez:2005cj,Shibata:2005mz,Stephens:2006cn,Stephens:2008hu} 
and three-dimensional simulations~\cite{Shibata:1999yx,Duez:2004uh,Baiotti:2004wn,Baiotti:2007np,Giacomazzo:2011cv,Reisswig:2012nc}.   
The relevant theoretical questions are related to the nature of the
collapse spacetime, black hole formation and its properties.
Astrophysically, rotating neutron stars close to
the collapse threshold can be produced in stellar core collapse or
neutron star mergers,
e.g.~\cite{Ott:2010gv,Thierfelder:2011yi}. Associated to such events,
a significant emission of electromagnetic, neutrino and gravitational
radiation is expected, e.g.~\cite{Andersson:2013mrx}. Accurate numerical
relativity simulations are essential to develop emission models.
Thus, understanding the technical details of the simulations, such as the
role of the gauge and the sources of inaccuracies, is of particular importance. 
In this paper we reexamine two key aspects of the rotating collapse by
a new set of numerical relativity simulations. 

First, we investigate the end state of the collapsing spacetime when
puncture gauge conditions are adopted, and compare it to the spacetime
of a single spinning puncture. 
Gauge conditions are a key technical point for the simulation of
collapse in 3+1 general relativity. The combination of the 1+log
slicing condition~\cite{BonMasSei94} and Gamma-driver shift for the
spatial gauge~\cite{AlcBruDie02,MetBakKop06}, commonly referred as
``puncture gauge''~\footnote{
The puncture gauge is effectively used within the moving puncture
approach introduced in~\cite{Campanelli:2005dd,Baker:2005vv}.},       
allows one to perform stable simulations                                  
and follow black hole formation without excision 
treatment~\cite{Baiotti:2006wm}. A clear understanding of the role of
this gauge in the gravitational collapse scenario has been achieved
only in the spherically symmetric case~\cite{Thierfelder:2010dv}. 
For axisymmetric spacetimes little is known. In vacuum, it is unclear
how and to what stationary slice of Kerr the conformally flat spinning
puncture initial data evolve. Some numerical and analytical studies
have recently been performed in~\cite{DieBru14,DenBauPed14}. 
Here, we present the first analysis in presence of matter.

Second, we calculate the gravitational waveforms (GWs) emitted during
collapse. Consistent gravitational waveforms from the neutron star
collapse can be computed only using full general relativity.
It has been pointed out long ago, and notably
in~\cite{Stark:1985da,Seidel:1987in,Seidel:1988za,Seidel:1990xb}, that
the relevant features are rather simple and waveforms resemble the
ones generated by a particle infalling the 
black hole~\cite{Davis:1971gg,Nakamura:1987zz}.
However, several three-dimensional studies suggest a more complicated wave pattern with the
exception of recent work of~\cite{Reisswig:2012nc} (see
also~\cite{Giacomazzo:2012bw}) in which the 
``perturbative picture'' holds. Our data confirm the latter result.
The investigation of these aspects requires very precise numerical
data. In this work, such precision is achieved by the use of (i) a
conservative mesh refinement scheme for the hydrodynamics
evolution~\cite{Berger:1989,East:2011aa,Reisswig:2012nc}, and (ii) the
Z4c formulation of Einstein equations~\cite{Bernuzzi:2009ex}, which is
applied for the first time to this problem.  

The paper is organized as follows. Section~\ref{sec:met} summarizes the
equations, the numerical method, and the implementation details. 
Section~\ref{sec:dyn} presents the dynamics of the gravitational collapse.
Section~\ref{sec:bh} compares our numerical results with the spacetime 
of a spinning puncture.
Section~\ref{sec:wav} focuses on the emitted gravitational waves. 
We conclude in Sec.~\ref{sec:conc}.
Throughout the article dimensionless units are used, i.e.~we set 
$c=G=M_\odot=1$.

\section{Method}
\label{sec:met}

\subsection{Numerical relativity framework}
\label{sec:met:gen}

Let us start discussing briefly the general relativity framework employed in
this work. Einstein's field equations are written in 3+1 form and
formulated as the Z4c system~\cite{Bernuzzi:2009ex,Hilditch:2012fp}.
The gauge conditions are specified as evolution equations for the 
lapse function $\alpha$ and the shift vector $\beta^i$. 
We employ the 1+log slicing condition~\cite{BonMasSei94}, 
\be
\label{alphadot}
\p_t \alpha = \beta^i \p_i \alpha -\alpha^2 \mu_L \hat{K} \ ,
\ee
together with the integrated version of the Gamma-driver shift 
condition~\cite{AlcBruDie02,MetBakKop06}
\be
\label{betadoti}
\p_t \beta^i = \mu_S \Gamma^i - \eta \beta^i + \beta^j \p_j \beta^i \ .
\ee
Above, $\hat{K}$ is the trace of the extrinsic curvature in the Z4c
formulation and $\Gamma^i$ the conformal connection functions. 
These conditions are commonly referred to as ``puncture
gauge''. Puncture gauge conditions have been proved to be a key
element for collapse simulations~\cite{Baiotti:2006wm,Thierfelder:2010dv}. 
The gauge parameters in Eqs.~(\ref{alphadot})-(\ref{betadoti}) are
chosen as $\eta=0.3$, $\mu_L=2/\alpha$, and $\mu_S=1$.
We employ the constraint damping scheme of the Z4c formulation, and
set the damping parameters to $\kappa_1=0.02$ and
$\kappa_2=0$~\cite{Weyhausen:2011cg}. 

The neutron star matter is described within general relativistic
hydrodynamics (GRHD)~\cite{Font:2007zz}. Eulerian GRHD equations are written
in conservative form and coupled with the evolution equations for the
spacetime. We use the same notation and equations as described
in~\cite{Thierfelder:2011yi}, and refer to that paper for details.
The equation of state for the fluid considered here is the
$\Gamma$-law, 
\be
\label{gammalaw}
p =(\Gamma-1) \rho \epsilon \ , 
\ee
where $p$ is the fluid pressure, $\rho$ the rest-mass density,
$\epsilon$ the specific internal energy, and $\Gamma$ the adiabatic
exponent.

\subsection{BAM code}
\label{sec:met:bam}

For our simulations we use the BAM code described
in~\cite{Thierfelder:2011yi,Brugmann:2008zz}. The numerical method is
based on the method of lines, Cartesian grids and finite
differencing. BAM implements a grid made of a hierarchy of
cell-centered nested Cartesian boxes. The grid structure is build out of $L$
levels of refinement labeled~$l = 0,...,L-1$. Every refinement level
$l$ has one or more Cartesian grids with constant grid spacing $h_l$
and $n$ points per direction. Levels are typically refined in
resolutions of constant factors of two. Levels with $l>l_{\rm m}$ can
employ a different number of points per direction, $n_{\rm m} \neq n$.
Runge-Kutta type integrators are used for the time evolution. 
For the time stepping  the Berger-Oliger algorithm (BO) is 
employed~\cite{Berger:1984zza}.
Metric spatial derivatives are approximated by fourth-order finite
differences. GRHD equations are solved with a standard
high-resolution-shock-capturing scheme based on primitive reconstruction
and the Local-Lax-Friedrich central scheme for the numerical
fluxes. Primitive reconstruction is performed with the fifth-order 
weighted essentially non-oscillatory (WENO)
scheme of~\cite{Borges20083191} (see~\cite{Bernuzzi:2012ci} for its
application in numerical relativity).

The main difference with respect to previous work is the implementation of an
algorithm to enforce mass conservation of the hydrodynamical quantities among
different refinement levels~\cite{Berger:1989} (see
also~\cite{East:2011aa,Reisswig:2012nc} for numerical relativity
implementations).  
This algorithm allows us to
use refinement levels inside the neutron star without introducing 
mass violation. 
Our implementation follows~\cite{East:2011aa};
details and extensive tests for single and binary neutron star spacetimes
will be given elsewhere~\cite{DieBerUje15}. We mention 
that, throughout this work, we use averages for the BO restriction and 
a second-order ENO (essentially non-oscillatory) scheme for the BO prolongation step.

Simulations presented in this work employ quadrant symmetry. 
The grid configurations are described in Table~\ref{tab:gridBAM}. 
We investigate numerical uncertainties by increasing both the number
of refinement levels keeping the same number of points per directions, and
the resolution for a fixed number of levels. The former procedure
allows us to better resolve the origin and the puncture in an efficient
way; the latter has usually a larger effect on the
waveforms. In the next sections these effects are discussed.

\begin{table}[t]
  \centering    
  \caption{Grid configurations for the BAM simulations: 
    $L$ number of total levels, 
    $n$ number of points per direction, $L_{ \rm m}$ number levels employing
    $n_{\rm m}$ points per direction (every level $l>3$), 
    $h_f$ finest grid spacing, $h_c$ coarsest grid spacing. 
    The neutron star is covered completely by level $l=5$, while its equatorial radius is $\sim7.7M_\odot$.    
    The outer boundary is roughly at $r_{\rm out}\sim576M_\odot$.}
  \begin{tabular}{l|ccccccc}        
    \hline
    Name & $L$ &  $n$ & $L_{\rm m} $ & $n_{\rm m}$ & $h_f$ & $h_c$ \\
    \hline
    G8    & 8   & 144 & 4 & 64 & $0.0625$    & $8$     \\
    \hline
    G9L  & 9   & 108 & 5 & 48 & $0.04167$   & $10.67$ \\
    G9   & 9   & 144 & 5 & 64 & $0.03125$   & $8$     \\
    G9H  & 9   & 216 & 5 & 80 & $0.025$     & $6.4$   \\
    \hline
    G10   & 10  & 144 & 6 & 64 & $0.015625$  & $8$     \\
    \hline
    G11  & 11  & 144 & 7 & 64 & $0.0078125$ & $8$     \\
    G11H & 11  & 216 & 7 & 96 & $0.0052083$ & $5.33$     \\
    G11F & 11  & 288 & 7 & 128& $0.00390625$ & $4$     \\
    \hline
  \end{tabular}
  \label{tab:gridBAM}
\end{table}

\section{Collapse dynamics}
\label{sec:dyn}

\begin{figure}[t]
\begin{center}
 \includegraphics[width=0.5\textwidth]{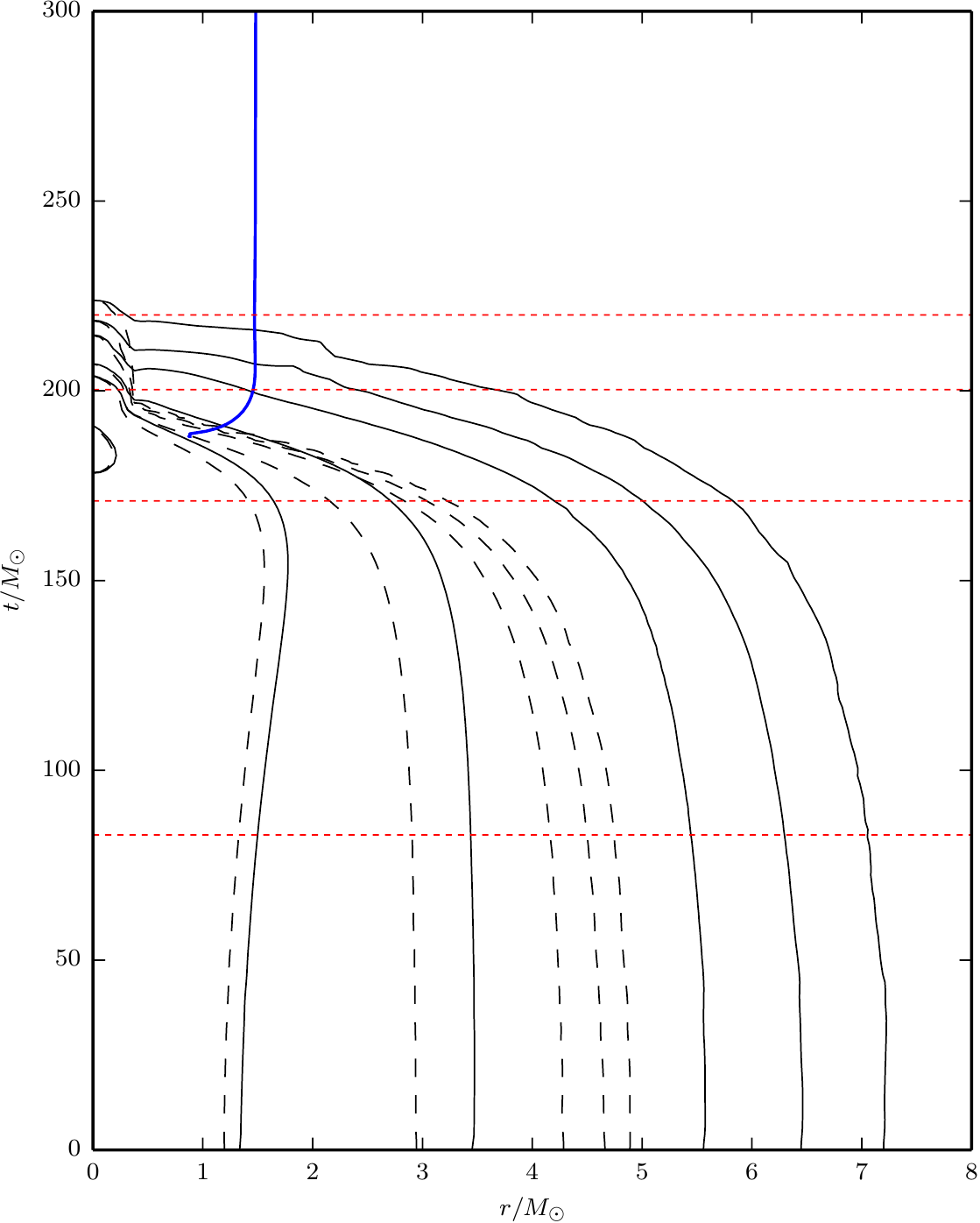}
\caption{Spacetime diagram visualizing the collapse dynamics of G11H. 
Contour lines in the equatorial plane (black solid) and perpendicular plane
(black dashed) are shown for $\rho = 2.5\cdot 10^{-5},10^{-4},
2.5 \cdot 10^{-4}, 10^{-3}, 2.5\cdot 10^{-3}, 10^{-2}$.
The apparent horizon forms at $188M_\odot$ (straight blue line).
Red dashed horizontal lines correspond to special features of in the
gravitational wave signal as marked in Fig.~\ref{Fig:wav}.}        
  \label{Fig:spacetime_diag}
\end{center}
\end{figure}

We study the rotational collapse by evolving a particular initial
stellar configuration constructed by perturbing a uniformly rotating
neutron star model in unstable equilibrium, i.e.~beyond the radial
stability point. In this section we describe the dynamics of the
collapse and the grid/resolution dependence in our simulations. 

The initial data configuration is the D4 model investigated previously
in~\cite{Baiotti:2004wn,Baiotti:2007np,Giacomazzo:2012bw,Reisswig:2012nc}. This
choice facilitates the comparison with previous work. The equation of
state is a $\Gamma=2$ polytrope $p= K\rho^\Gamma$ with $K=K_{\rm ID}=100$,
the model's central rest-mass density is $\rho_c = 4.0869 \cdot 10^{-3}$, 
the axes ratio is $0.65$, the gravitational mass $M=1.8605 M_\odot$, and the
baryonic mass $M_b=2.0443 M_\odot$.  
The equilibrium configuration has been computed with Stergioulas's RNS
code~\cite{Stergioulas:1994ea}. 

The gravitational collapse can be induced either by a pressure
perturbation or by computing initial data at low resolution. 
Both methods violate Einstein constraints; the violation can affect 
significantly the calculation of gravitational radiation. 
For large perturbations or low-resolution initial data we observe a
large unphysical burst of radiation at early times; in some cases this
burst has an amplitude comparable to the waveform amplitude and cannot
be clearly separated from the physical data.
We minimize this effect by choosing a small perturbation and using
high resolutions. Specifically, we reduce the initial pressure of $0.5\%$
by recomputing its equilibrium value with a different polytropic
constant, $K=99.5$, then the unperturbed one $K_{\rm ID}=100$ (compare
with~\cite{Baiotti:2004wn,Baiotti:2007np,Reisswig:2012nc} where $2\%$
and~\cite{Giacomazzo:2012bw} where $0.1\%$ was applied). The model is then evolved with Eq.~(\ref{gammalaw}).

The collapse dynamics is summarized by the simplified spacetime
diagrams shown in Fig.~\ref{Fig:spacetime_diag}, which shows the
evolution of the coordinate star surface, constant density lines, 
and apparent horizon radius; see
e.g.~\cite{Baiotti:2004wn}. Most of the matter contracts in an almost
homologous way and maintains its axisymmetric distribution until
$t\sim175M_\odot$. Notice, however, that at high densities
($r\lesssim2$) the contour lines slightly expand before collapsing.
An apparent horizon is first found at $t\sim188M_\odot$ (for resolution G11H). 
Soon after horizon formation, all the matter is inside the horizon and
actually ``falls  off'' the grid due to gauge conditions
(see~\cite{Thierfelder:2010dv} 
and below). In Sec.~\ref{sec:wav} we will further discuss this
spacetime diagram, and identify specific waveform features for each
time marked in Fig.~\ref{Fig:spacetime_diag}.

\begin{figure}[t]
\begin{center}
 \includegraphics[width=0.5\textwidth]{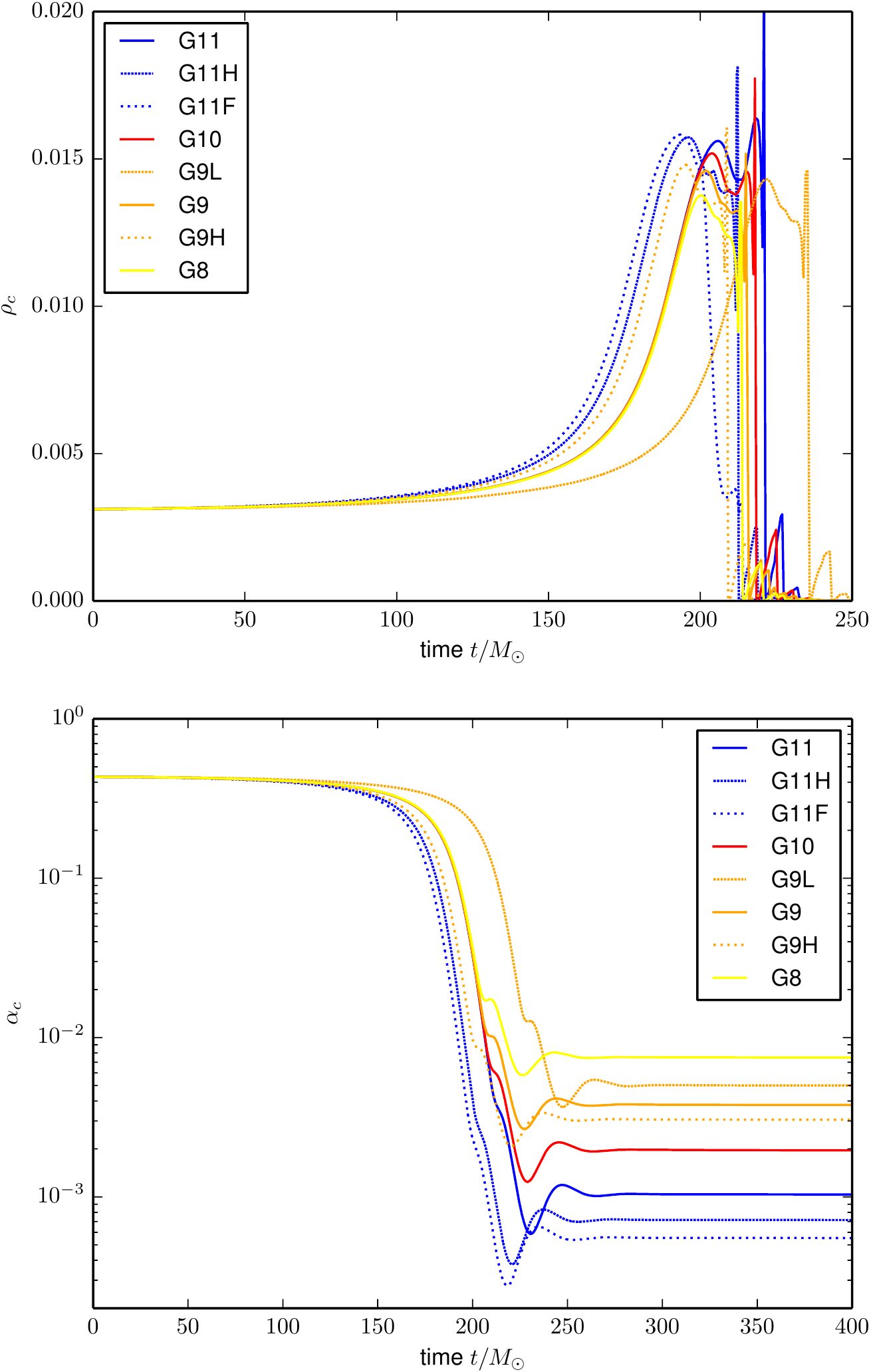}
 \caption{Rotating collapse central dynamics. Central rest-mass
 density (top) and central lapse (bottom). Results for various grid
 configurations and resolution are shown.} 
  \label{Fig:D4_dyn_central}
\end{center}  
\end{figure}

Figure~\ref{Fig:D4_dyn_central} shows the evolutions of the
central (coordinate radius $r\approx0$) density $\rho_c$ and the central
lapse $\alpha_c$. During collapse the central lapse decreases and the
central density increases; the latter reaches a maximum  
at $t \sim 195 M_\odot$.
The plot shows results for different grid configurations. By increasing
the number of refinement levels the origin is better resolved, and
consequently the maximum density (lapse) increases (decreases). Notice
this is consistent with the argument of~\cite{Thierfelder:2010dv}.   
By varying the resolution for a given grid setup we observe a
monotonic behavior. The resolution effect (see the dashed-solid-dotted
lines for configurations G9 and G11) can be as large as the effect of
including more refinement levels; hence both parameters need to be
tuned for an optimal grid setup. For higher resolutions the collapse
happens earlier. 

\begin{figure}[t]
\begin{center}
 \includegraphics[width=0.48\textwidth]{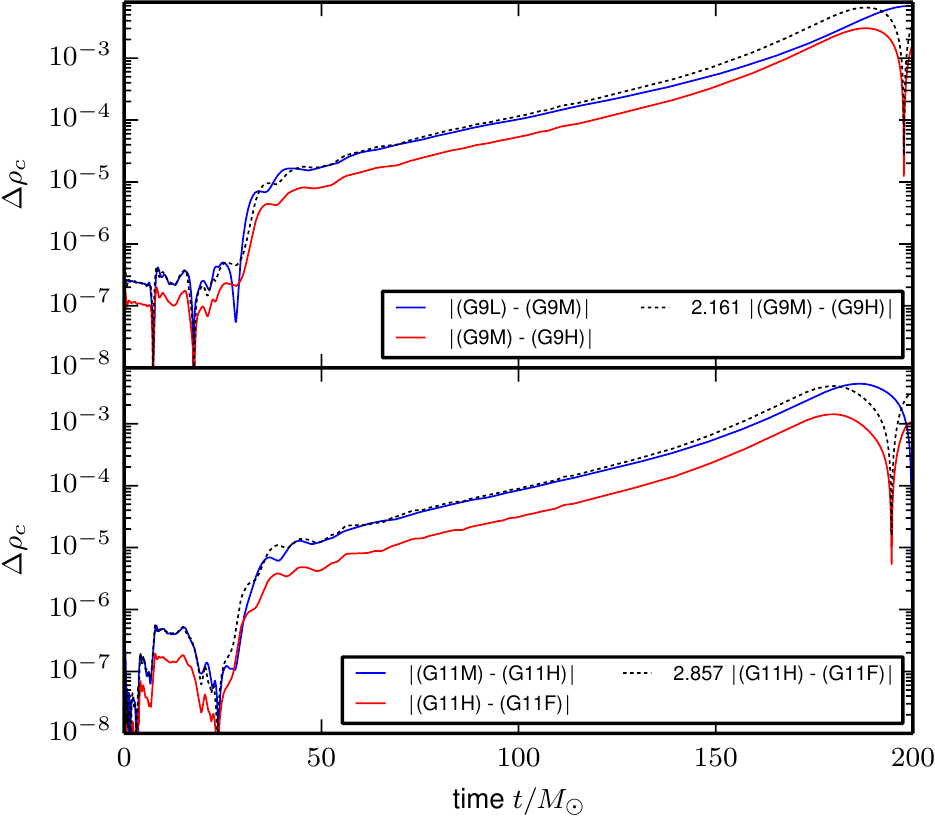}
 \caption{Convergence test for the density $\rho_c$ for the G9
 (upper panel) and G11 (lower panel) grid configurations. For both   
 triplets the results scale at approximately second order; convergence is more
 robust and observed longer for the higher resolved G11 data.}   
  \label{Fig:D4_dyn_conv}
\end{center}
\end{figure}

In Fig.~\ref{Fig:D4_dyn_conv} we show a standard three-level
self-convergence plot for the central density $\rho_c(t)$. Similar
plots are also obtained for other quantities. For the G11 configurations we
observe second-order convergence almost up to horizon formation (see
later), while for G9 convergence is slower after $t\sim100M_\odot$. After
horizon formation convergence is slower, and, in particular, cannot be
monitored \textit{at} the origin, when the black hole (puncture) forms.
In order to minimize the numerical uncertainty for our local analysis
presented below, we use the highest number of refinement levels  
and focus on the model G11H (unless otherwise stated).

\begin{figure}[t]
\begin{center}
 \includegraphics[width=0.5\textwidth]{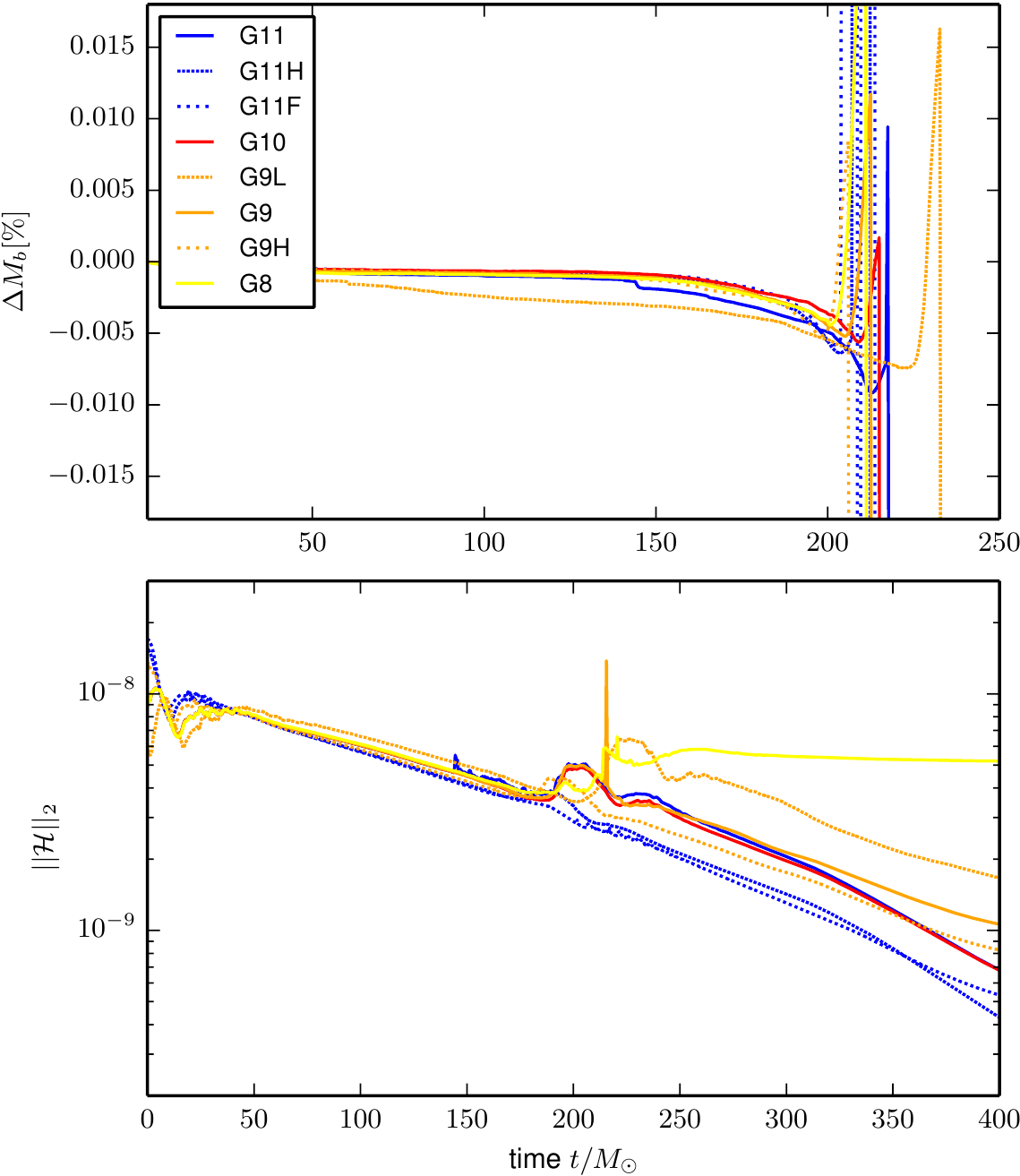}
 \caption{Rotating collapse dynamics and global quantities: conservation
 of baryonic mass $\Delta M_b=1-M_b(t)/M_b(0)$ (top), $L_2$ norm of the
 Hamiltonian constraint 
 $||\mathcal{H}||_2$ (bottom). The
 constraint violations are measured on level $l=1$, i.e. the second
 coarsest level on which the wave extraction also takes place.}   
  \label{Fig:D4_dyn_constraint}
\end{center}  
\end{figure}

Fig.~\ref{Fig:D4_dyn_constraint} shows the baryonic mass
conservation, and the $L_2$ norm of the Hamiltonian constraint
$||\mathcal{H}||_2$. 
The relative error in the mass conservation is $\lesssim10^{-4}$
up to the collapse. At black hole formation the Hamiltonian
constraint (and the momentum constraints, not in the figure)
shows a maximum. Constraint violations decrease when the grid 
is refined and the origin better resolved. Additionally, the higher the
resolution is, the smaller the constraint violations are. 
Notably, for the higher resolutions the violation remains below the
level of the initial data due to the use of the Z4c formulation.  
 
\begin{figure}[t]
\begin{center}
 \includegraphics[width=0.48\textwidth]{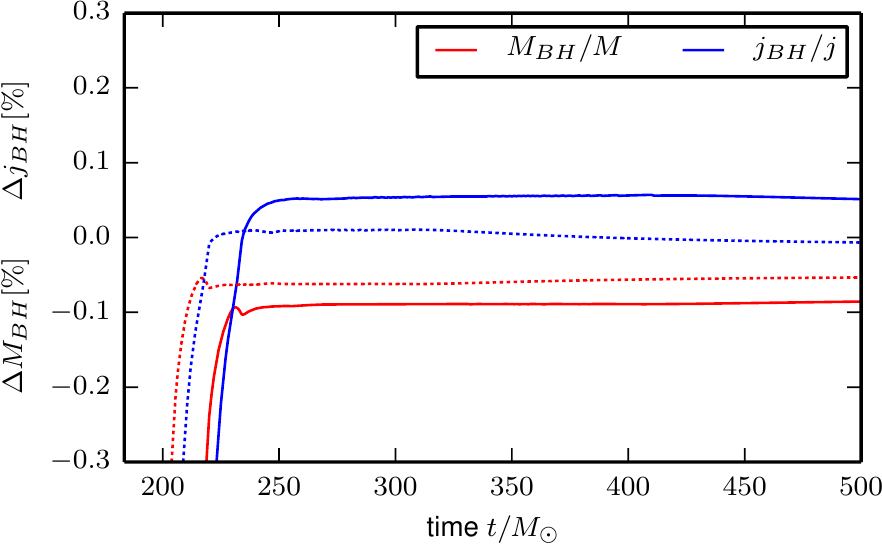}
 \caption{Differences between the horizon mass (red) and dimensionless
  angular momentum (blue) of the final black hole corrected by the
 radiation with respect to the initial ADM quantities of the star.
 For both quantities the error is below $0.1\%$. 
 G10 grid data are shown with solid lines; dashed lines are used for
 G11F data.}   
  \label{Fig:horizon}
\end{center}
\end{figure}

The horizon mass and angular momentum, as measured by the apparent
horizon finder, are $M_{\rm BH}\sim  1.859(1) M_\odot$ and $j_{\rm  BH}\sim0.543(7)$. 
In Fig.~\ref{Fig:horizon} we show the differences between the horizon
mass and spin with respect to the initial ADM quantities and those
estimated by the apparent horizon corrected by the
amount of energy (angular momentum) emitted in gravitational waves
(see below). We find typical relative errors at, or below, the $\sim
0.1\%$ level.

\section{Collapse  end state}
\label{sec:bh}

In vacuum, the numerical evolution of puncture black hole
initial data~\cite{Brandt:1997tf} approaches an asymptotically
cylindrical stationary solution called \emph{trumpet}~\cite{Hannam:2006vv}. The spatial gauge
choice, in particular, is responsible for pushing grid points close to
the puncture into the black hole interior; the initial wormhole
topology ceases to be numerically
resolved~\cite{Brown:2007tb,Garfinkle:2007yt,Hannam:2008sg}. 
The end state of a spherical gravitational collapse asymptotically approaches
the same trumpet solution found 
in vacuum simulations~\cite{Thierfelder:2010dv}. The agreement of
end states is again caused by the spatial gauge condition, which
allows the matter to fall inwards into a region of spacetime that is
not resolved by the numerical grid. As stressed
in~\cite{Thierfelder:2010dv},  the result is nontrivial because in the
collapsing spacetime there is (at least in the matter region) no
timelike Killing vector that can lead to a stationary end state, and,
at the continuum level, it has different topology than the puncture. 
Trumpet solutions are also found in dust and gravitational wave
collapse~\cite{Staley:2011ss,HilBauWey13}.  
In axisymmetric vacuum spacetimes, one can argue that puncture initial
data evolve towards some stationary trumpet slices of
Kerr~\cite{DieBru14,DenBauPed14,Hannam:2009ib,GabachClement:2009dk,Dain:2008yu}. In~\cite{DieBru14}
a first numerical examination of spinning black holes with the  
puncture gauge was performed and, recently, Ref.~\cite{DenBauPed14}
found an analytical description of particular trumpet slices in the
Kerr spacetime. Our discussion builds on the results
of~\cite{DieBru14}. 

\begin{figure}[t]
\begin{center}
  \includegraphics[width=0.48\textwidth]{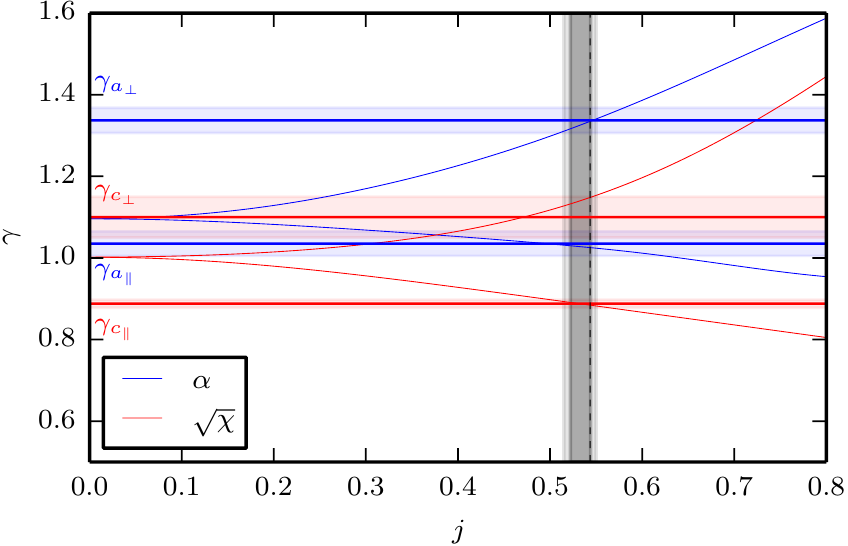}
  \caption{Characteristic behavior of metric variables $\sqrt{\chi}$
    and $\alpha$ at $r=0$. Exponents $\gamma_c$ and
  $\gamma_{a,c}$ are extracted by fitting to Eq.~(\ref{eq:chifit}) for G11H. The
  thin solid lines are spinning puncture data~\cite{DieBru14}. The thick lines are collapse
  data with error bars obtained from our simulations. Red (blue) color
  refers to $\chi$ ($\alpha$). The dashed vertical line indicates the
  angular momentum from the apparent horizon finder. 
  The dark shaded region represents the spin obtained from Eq.~\ref{eq:j_K}, 
  the light shaded region the estimate according to $\gamma_{a,c}$.} 
  \label{Fig:chialpha}
\end{center}
\end{figure}

In the following we demonstrate that the end state of a rotating, collapsing 
neutron star is a spinning puncture of mass $M$. 
We propose two arguments for this statement; both arguments rely on
the fact that various metric functions {\it at the puncture} 
can deliver information about the puncture's spin~\cite{DieBru14}. In
particular, the leading-order behavior of the (square root of the)
conformal factor and of the lapse function is 
\be
\sqrt{\chi} (r\sim0)  \sim c_0  +  c_1 r^{\gamma_c}, \ \mbox{and} \ 
\alpha      (r\sim0) \sim a_0  +  a_1 r^{\gamma_a} \ ,
\ee
with $\gamma_c$ and $\gamma_a$ characteristic exponents that depend on
the spin (see Fig.~2 of~\cite{DieBru14} and Fig.~\ref{Fig:chialpha} below).
Furthermore, the dimensionless spin $j$ of a puncture can be estimated as
\be
\label{eq:j_K} 
 j \simeq \sqrt{1.41789-4.71218 \cdot \bar{K}(r=0)} \ ,
\ee
extracting the value of the extrinsic curvature, $\bar{K}=\hat{K}/M$, at
point $r=0$. In the following, we verify that the spin 
estimated in the collapsed spacetime from $\gamma_c$, $\gamma_a$, and
$\bar{K}(r=0)$ agrees with the angular momentum measured from the
apparent horizon.

The exponents $\gamma_c$ and $\gamma_a$ can be determined as best fits of
the simulation data according to the models, e.g.
\be
\label{eq:chifit} 
 \sqrt{\chi}(r\sim0) = c_0  +  c_1 r^{\gamma_c}  (1+c_2 r +c_3 r^2) 
\ee
and similarly for the $\alpha (r\sim0)$. The fit is calculated on the
radial interval $r\in[0.01,0.3]$ in a direction either
parallel or perpendicular to the 
rotational axis ($z$-axis). Note that the parallel and perpendicular values actually
differ~\cite{DieBru14}. The results are reported in
Fig.~\ref{Fig:chialpha}. The thin solid lines are spinning
puncture data in the parallel and perpendicular direction. 
The straight thick lines are collapse data with error bars estimated with 
the help of different resolutions and different fitting intervals for $r$.
Red (blue) color refers to $\chi$ ($\alpha$). The vertical line indicates
the dimensionless angular momentum estimated from the collapse
simulation's apparent horizon. The figure shows that the spinning
puncture lines cross the collapse data at these points.  
The dimensionless angular momentum is compatible with the one of
a puncture of the same mass. 

Let us also consider a second estimate of the dimensionless spin based on the 
evolution variables and the puncture gauge. According
to~\cite{DieBru14} the extrinsic curvature depends on the angular
momentum of the black hole,  when a stationary state is reached. 
The value $\bar{K}(r=0)= \hat{K}(r=0) \cdot M$ can be extrapolated from a linear
fit of $\hat{K}$ in the  region $r \in [0.05;0.25]$. 
We use an extrapolation perpendicular, orthogonal, and in
an angle of $45^\circ$ to the spin axis.  
In principle all these values coincide~\cite{DieBru14}.  
We receive $K(r=0)=0.1301$ along the $x$-axis, $K(r=0)=0.1284$ along
the $z$-axis, and $K(r=0)=0.1293$ along the diagonal for the G11H setup. Using Eq.~(\ref{eq:j_K})
we get $j=0.533\pm 0.014 $, which agrees with the measured horizon
spin within the measurement uncertainty (which is obtained from different resolutions and fitting intervals).

\section{Gravitational waveforms}
\label{sec:wav}

 \begin{figure*}[t]
 \begin{center}
  \includegraphics[width=1\textwidth]{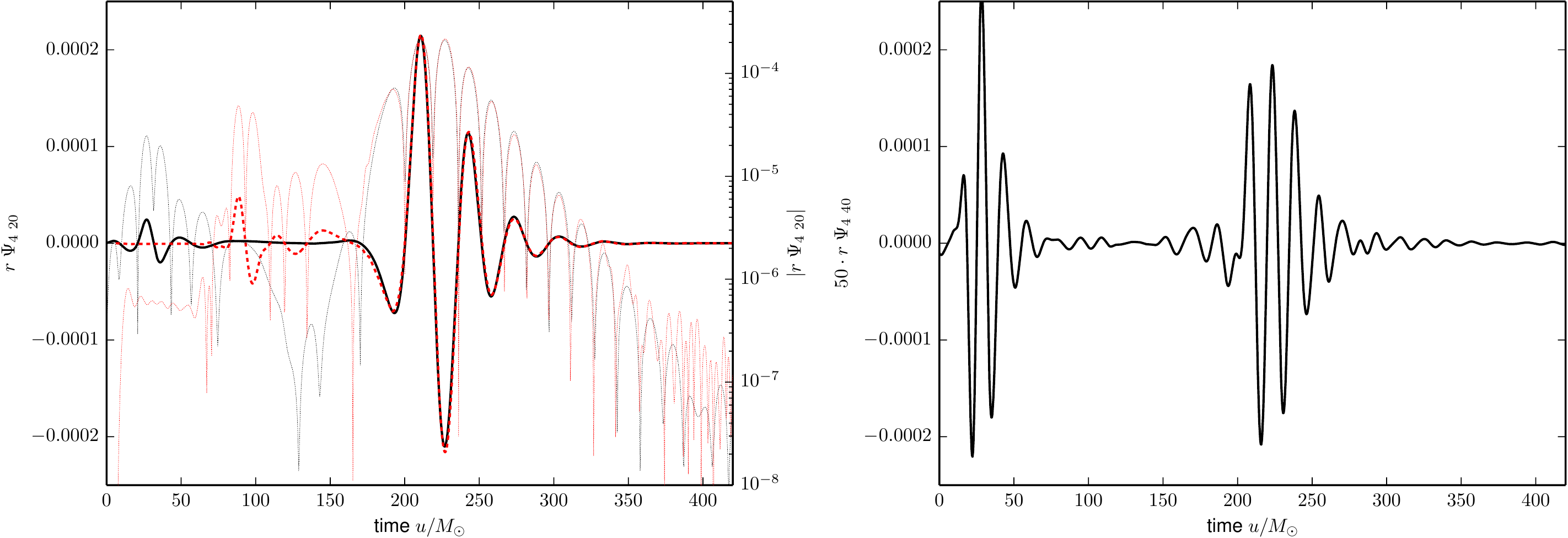}
  \caption{Rotating collapse curvature waveform $r\Psi_{4 \ \ell m}$
  for the dominant multipoles $(\ell,m)=(2,0),(4,0)$. Left: The
  $(\ell,m)=(2,0)$ mode is shown in linear scale (solid lines) and its
  modulus in log scale (dotted lines) to highlight the ringdown.
  The waveform of~\cite{Reisswig:2012nc}
  is shown in red for comparison.
  Right: The $(\ell,m)=(4,0)$ mode is not well resolved by the simulation.} 
   \label{Fig:wav_psi}
 \end{center}
 \end{figure*}

Gravitational waves are computed by multipole decomposition of the
curvature invariant $r\Psi_4$; metric multipoles $rh_{\ell m}$ are
then reconstructed from curvature multipoles (see below).
In the following, we discuss both curvature and metric waveforms.
Most of the GW energy $E_{gw}\sim 7.5 \cdot 10^{-7}$ is emitted in the
$(\ell,m)=(2,0)$ channel. The 
second dominant mode is the $(\ell,m)=(4,0)$ multipole, but, as we
shall see, it cannot be computed accurately.  
We plot waveforms against a retarded time defined as
$u=t-r_*=t-r-2M\log\left(r/2M-1\right)$.

Figure~\ref{Fig:wav_psi} shows the two dominant axisymmetric modes
$\ell=2,4$ of the curvature waveform. The left panel plots the
quadrupolar $\ell=2$ mode, which is characterized by a burst of
radiation peaking \emph{before} black hole formation and followed by a
ringdown pattern.  We also show the $|\Psi_{4_{(20)}}|$ in log scale
to highlight the quasinormal ringing phase. 
According to the ten local maxima between $u\in[225
M_\odot,380 M_\odot]$, we calculate the fundamental complex
frequencies and find $M\omega=(0.425,-0.0842)$. 
Comparing our results with~\cite{Berti:2009kk} we see that our values
agree within $(10\%,3\%)$ with perturbation theory~\cite{Berti:2009kk}
assuming $j=j_{\rm BH}=0.544$ and stationarity. (Notice the spacetime
in the simulation is still very dynamical at $t\sim225M_\odot$.)   

The left panel of Fig.~\ref{Fig:wav_psi} compares our data with those
of~\cite{Reisswig:2012nc}, extracted at scri and kindly provided by
the authors. Waveforms are shifted in time to match the peaks. The
comparison indicates a very good agreement~\footnote{%
The $\Psi_{4\ 20}$ waveform visually agrees also
with the one of~\cite{Giacomazzo:2012bw}.}. 
Notice that, as in our work,~\cite{Reisswig:2012nc} also uses a
conservative mesh refinement algorithm, but employs the BSSN-evolution 
system~\cite{NakOohKoj87,ShiNak95,BauSha98} 
and wave extraction is performed with the Cauchy-characteristic extraction. 

As one can observe from the figure's right panel, the $\ell=4$-mode
has amplitude $\sim50$ times smaller then the $\ell=2$. The amplitude
is of the same order as the burst of radiation caused by 
the initial (constraint violating) perturbation at early times. 
These kinds of data are inaccurate, and should be
discarded in a physically meaningful analysis.

\begin{figure}[t]
\begin{center}
 \includegraphics[width=0.5\textwidth]{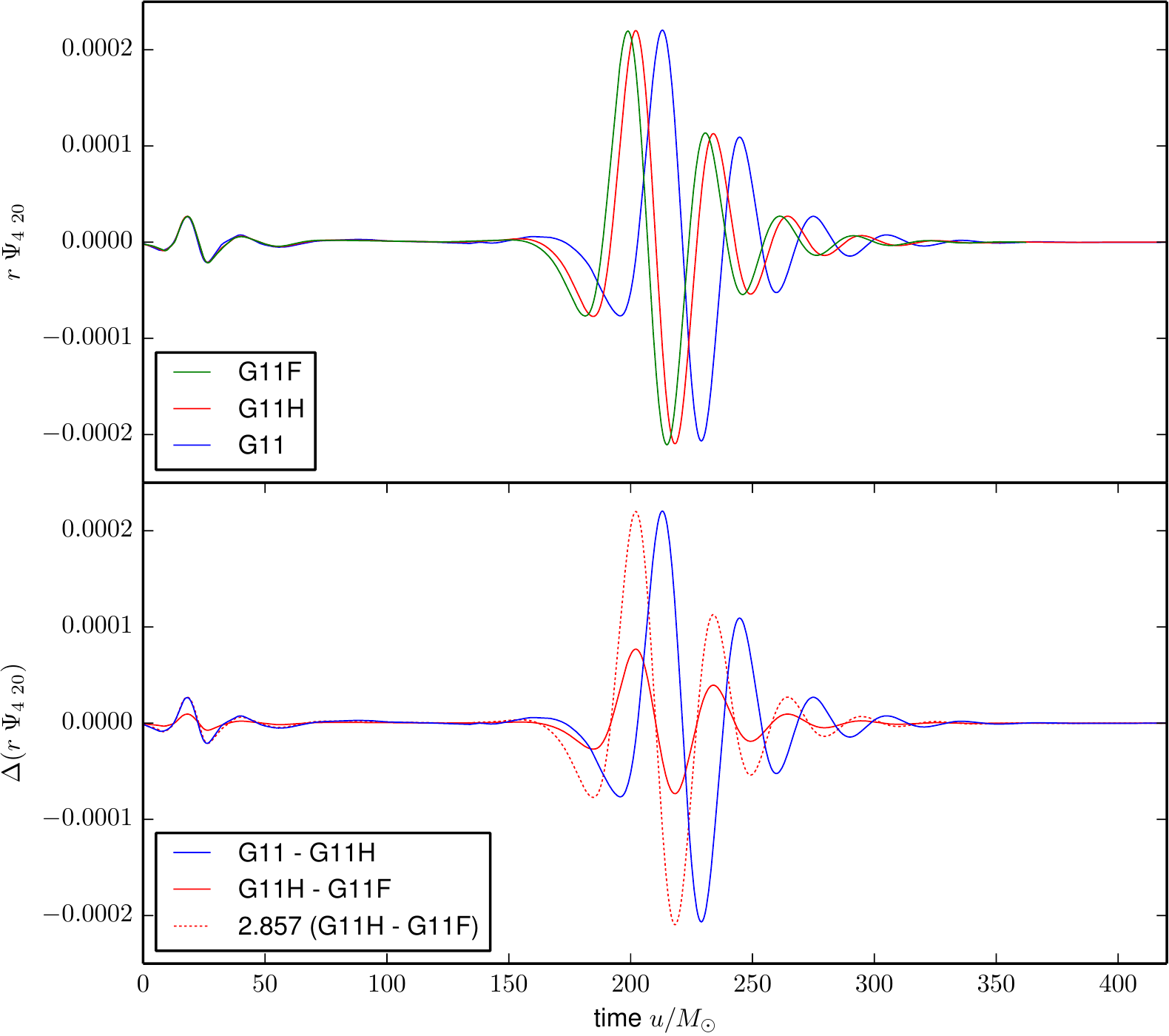}
 \caption{Rotating collapse curvature waveform $r\Psi_{4 \ 20}$,
 convergence study.} 
 \label{Fig:wav_psi_conv}
\end{center}
\end{figure}

A self-convergence test on the $(\ell,m)=(2,0)$ waveform is shown in
Fig.~\ref{Fig:wav_psi_conv}. 
We observe approximate second-order convergence in
the G11 data. However, clear pointwise convergence of the waveform is difficult
to obtain: since we evolve constraint violating initial data,
simulations at different resolutions are inconsistent, and for
instance, they do not tend to the same continuum collapse
time. Although the effect is rather small, it is visible in the
convergence plot as a dephasing in the differences.
The effect is larger at lower resolutions (and for larger initial
perturbations, not discussed here), but persists also at high
resolutions. We expect it can be removed only by using constraint
satisfying initial data. 

Further, we study uncertainties due to finite radius extraction. 
Waveforms computed at different radii $r=(100,150,200,250,300)$ and
plotted against $u$ slightly differ in amplitude. A linear
extrapolation to $r\to\infty$ of $r\Psi_{4\ 20}(u;r)$ shows that the
amplitude uncertainty can be as large as $15\%$ for $r=100$ and drop
to below $5\%$ for $r=300$. This uncertainty can be of the same order
of truncation errors. Notice in this comparison the use of the
retarded time as defined above in terms of $r_*$ is essential in
order to properly align the waveforms, i.e.~the logarithm term
$2M\log\left(r/2M-1\right)$ has a significant contribution at these
radii. 

\begin{figure}[t]
\begin{center}
 \includegraphics[width=0.5\textwidth]{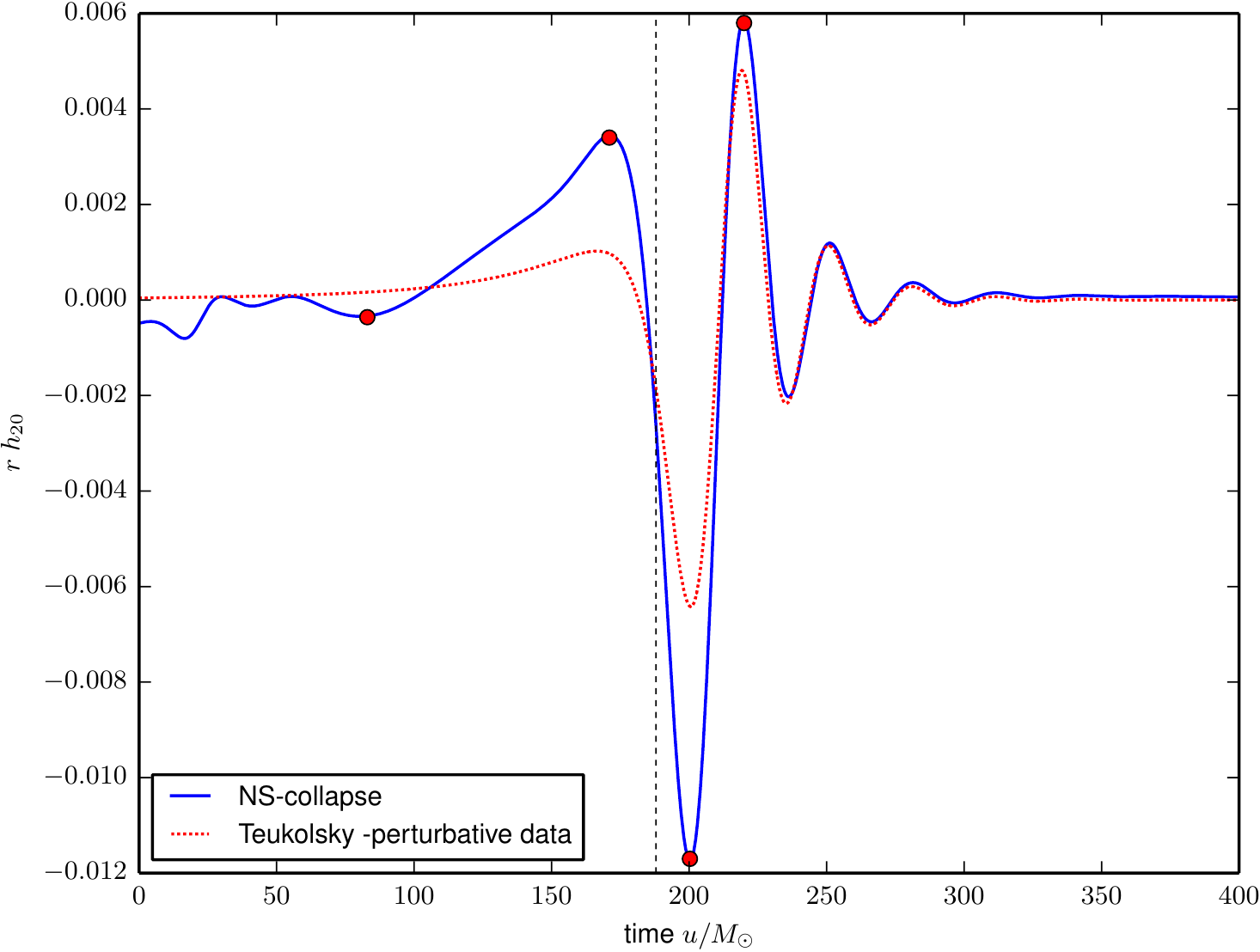}
 \caption{Rotating collapse metric waveforms $rh_{20}$. The $(2,0)$
 mode is compared with a Teukolsky perturbative simulation of black
 hole scattering; see text for details. 
 The waveform main features are marked with red dots and correspond to
 the events (horizontal lines) in Fig.~\ref{Fig:spacetime_diag}.
 The horizon formation is marked with a vertical dashed line.}  
  \label{Fig:wav}
\end{center}
\end{figure}

Let us turn now to the metric waveform, and discuss its physical
features. The multipoles $h_{\ell m}$ are reconstructed by integrating
the relation $\Psi_4=\ddot{h}$. We adopt a time domain integration
subtracting a quadratic polynomial as 
described in~\cite{Damour:2008te,Baiotti:2008nf}. Alternatively we
have experimented with the frequency domain integration
of~\cite{Reisswig:2010di}, but in the collapse problem it is difficult to
identify a cutting frequency for the high-pass filter proposed
there. In both cases the reconstruction introduces inaccuracies in the
ringdown.

The dominant mode of the metric waveform is shown in
Fig.~\ref{Fig:wav}.  
As pointed out
in~\cite{Stark:1985da,Seidel:1987in,Seidel:1988za,Seidel:1990xb}, 
the quadrupole waveform is particularly simple, and characterized by the
``precursor-burst-ringdown'' pattern well known from black hole
perturbation theory (either
scattering~\cite{Vishveshwara:1970zz,Bernuzzi:2008rq} or radially   
infalling
particles~\cite{Davis:1971gg,Davis:1972ud,Nakamura:1987zz,Lousto:1996sx}).    
The figure shows, together with our numerical
relativity calculation, the $\ell=2$ waveform obtained by a
perturbative Gaussian scattering experiment onto a Kerr black hole
with $j=j_{\rm BH}\sim0.544$~\cite{Harms:2014dqa}. The amplitude is
scaled by an arbitrary factor. The similarity of the numerical and
perturbative waveforms reflects the basic mechanism of the emission
process.   

It is interesting to connect the waveform features with the collapse
dynamics. In perturbation theory this is done, for instance, 
analyzing the background potential that drives the particle
motion~\cite{Davis:1971gg,Davis:1972ud}. For the
collapse dynamics of our study we use the spacetime
diagrams of Fig.~\ref{Fig:spacetime_diag} and connect the dynamics to
the emission using the retarded time $u=t-r_*$, i.e.~using null
geodesic of Schwarzschild spacetime. 
With these assumptions, the events marked in Fig.~\ref{Fig:spacetime_diag} with
horizontal lines correspond to the waveform features marked in
Fig.~\ref{Fig:wav}. The minimum in the precursor corresponds to time
$t\sim80M_\odot$, at which the collapse actually sets in. The first maximum
is related to the moment of time at which fluid particles significantly
accelerate, and is slightly antecedent apparent horizon formation. 
Indeed, we find that taking a worldline
$r(t)$ of Fig.~\ref{Fig:spacetime_diag}, the quadrupole waveform
$Q_{20}\propto \ddot{I}_{20}\propto - 2 \dot{r}^2 - 2 r \ddot{r}$,
captures all the qualitative features up to horizon formation.
The first maximum in particular is determined by the competitive
effect of the two terms in the quadrupole formula: $-\dot{r}^2<0$
and $-r\ddot{r}>0$. At times $t<150M_\odot$ the second term dominates,
$-r\ddot{r}>\dot{r}^2$, but at later times $t>150M_\odot$ the
first (velocity) term becomes comparable $\dot{r}^2\sim-r\ddot{r}$. 
The maximum in the wave at $t\sim175M_\odot$ results from the growth of
$\dot{r}^2$; the zero crossing at $t\sim180M_\odot$ marks the instantaneous
balance between the two terms.  
The metric waveform has its absolute minimum \textit{shortly after} black
hole formation (see dashed vertical line in Fig.~\ref{Fig:spacetime_diag}), 
when the mass enclosed by the horizon is $M_{\rm BH}\sim M$ and 
its radius is approximately constant. 
The metric waveform peaks \textit{after} black hole formation when all the
matter is inside the horizon and the black hole rings down.

\section{Summary}
\label{sec:conc}

Puncture gauge conditions play a key role in the simulations of
rotational collapse as they ``automatically'' handle the singularity
formation and subsequent
evolution~\cite{Baiotti:2006wm,Thierfelder:2010dv}. Building on
previous work and extending it, we have demonstrated that the end state of an
axisymmetric collapse in the puncture gauge is the same as the one
obtained from the evolution of a spinning puncture~\cite{DieBru14}. 
Our statement refers to a simple and controlled case study (an
unstable uniformly rotating equilibrium configuration perturbed to
collapse) but the result holds for generic simulations in which the
puncture gauge is employed.
For instance, rotational collapse characterizes the end phase of
certain binary neutron star configurations or supernova core
collapse. Not surprising, the same arguments used in 
this paper can be applied to those data, e.g.~\cite{DieBru14} for
preliminary results.

Our results strongly rely on the precision of the presented simulations. In
particular, we have used a conservative mesh refinement scheme for
the hydrodynamics
evolution~\cite{Berger:1989,East:2011aa,Reisswig:2012nc} which allowed
us to refine the star and increase the resolution near the center
(puncture) without mass losses. Also, we have employed the Z4c
formulation of Einstein equations, which improves accuracy and
constraint preservation in a free evolution (hyperbolic) approach to
general relativity~\cite{Bernuzzi:2009ex,Hilditch:2012fp}. 

The calculation of gravitational waves is particularly sensitive to
numerical resolution and errors. In these simulations, the waveform
quality can be corrupted by spurious radiation related to constraint
violations.  
Our data agree with the recent work of~\cite{Reisswig:2012nc}; some 
earlier three-dimensional calculations appear as affected by unphysical features
probably due to low resolution employed and the high initial perturbation. 
The collapse waveform is rather simple and qualitatively
similar (``precursor-burst-ringdown'') to those from black hole
perturbation theory~\cite{Davis:1971gg,Stark:1985da,Seidel:1990xb}. Using the
spacetime diagram of Fig.~\ref{Fig:spacetime_diag}, we have
identified and connected all its main features to precise stages of the
collapse dynamics.

\begin{acknowledgments}
  The authors thank Christian Reisswig and Christian Ott for
  discussions and for providing the waveform used in Fig.~\ref{Fig:wav_psi};
   Enno Harms for providing the data of the Teukolsky perturbative simulation used in
  Fig.~\ref{Fig:wav}; Bernd Br\"ugmann, Alessandro Nagar, Pedro Montero, and Nicolas
  Sanchis-Gual for discussions; and Nathan Johnson-McDaniel for comments on the manuscript.
    This work was supported in part by DFG grant SFB/Transregio~7
  ``Gravitational Wave Astronomy'' and the Graduierten-Akademie Jena.  
  S.B. acknowledges partial support from the National
  Science Foundation under Grants No. NSF AST-1333520,No. PHY-1404569, and
  No. AST-1205732.   
  The authors acknowledge the Gauss Centre for Supercomputing e.V. 
  for providing computing time on the GCS Supercomputer SuperMUC (Munich), 
  the John von Neumann Institute for Computing (NIC) providing
  computing time of JUROPA (JSC), and the Extreme Science and
  Engineering Discovery Environment (XSEDE) for providing computer
  time on Stampede (Texas).  
\end{acknowledgments}
\

\end{document}